\documentclass{actasen}

\begin{document}

\setcounter{page}{1}

\Volume{}{}


\runheading{Takumi MUTO}%

\title{Coexistence of Kaon Condensation and Hyperons in Hadronic Matter and Its Relevance to Quark Matter$^\dag$ }

\footnotetext{$^{\dag}$ Supported by the funds provided by Chiba Institute of Technology


\hspace*{5mm}$^{\bigtriangleup}$ takumi.muto@it-chiba.ac.jp\\

\noindent 0275-1062/01/\$-see front matter $\copyright$ 2011 Elsevier
Science B. V. All rights reserved. 
\noindent PII: }

\enauthor{Takumi MUTO$^{\bigtriangleup}$ }{Chiba Institute of Technology, 2-1-1 Shibazono, Narashino, Chiba 275-0023, Japan}
\enauthor{Toshiki MARUYAMA }{Advanced Science Research Center, Japan Atomic Energy Agency , Ibaraki 319-1195, Japan}
\enauthor{Toshitaka TATSUMI}{Department of Physics, Kyoto University, Kyoto 606-8502, Japan}

\abstract{Coexistence of kaon condensation and hyperons, which may be realized in neutron stars, is investigated on the basis of the relativistic mean-field theory combined with the effective chiral Lagrangian. 
It is shown that kaon-condensed phase in hyperon-mixed 
matter is plausible, but  it leads to significant softening of the equation of state (EOS).
We discuss indispensable effects which make the EOS stiffer so as to be consistent with recent neutron-star observations. 
}


\maketitle

\section{INTRODUCTION}

Various phases and equation of state (EOS) of dense matter in neutron stars have been elaborated actively through theoretical studies, terrestrial experiments and astronomical observations. 
Recently, observations of massive neutron stars as large as $2M_\odot$ ($M_\odot$ being the solar mass) have imposed a stringent constraint on the EOS that too soft EOS is excluded\rf{1}. 
On the other hand, theoretical studies have predicted that multi-strangeness systems such as hyperon ($Y$)-mixed matter and/or Bose-Einstein condensation of kaons ($K^-$) necessarily appear at high densities. In particular, coexistence of kaon condensation and hyperons in dense matter has been discussed in several approaches\rf{2}. However, most of the models including coexistent phase of kaon condensation and hyperon-mixed matter [($Y+K$) phase] lead to maximum neutron-star masses less than 1.85 $M_\odot$, which is not compatible with recent observations. 
We reconsider ($Y+K$) phase in neutron stars 
by the use of the interaction model based on the relativistic mean-field (RMF) theory for baryon ($B$)-$B$ interactions, coupled with the nonlinear effective chiral Lagrangian for $\bar K$-$B$ and $\bar K$-$\bar K$ interactions. We discuss what effects are necessary in order to reconcile theories with observations.  

\section{FORMULATION}

Within the RMF framework, the $B$-$B$ interactions are mediated by exchange of the $\sigma$, $\sigma^\ast$, $\omega$, $\rho$, and $\phi$ mesons. We take into account $\Lambda, \Sigma^-, \Xi^-$ for hyperons. The $\bar K$-$B$ and $\bar K$-$\bar K$ interactions are introduced from the nonlinear effective chiral Lagrangian\rf{3}, and 
the original $s$-wave scalar and vector contact interactions between the nonlinear $\bar K$ field and baryons are replaced by those generated by the scalar $\sigma$, $\sigma^\ast$ mesons-exchange and vector $\omega$, $\rho$, and $\phi$ mesons-exchange, respectively.  The classical $K^-$ field is taken to be the plane-wave type as
$ K\equiv\langle K^-|\hat K^-|K^-\rangle=f/\sqrt{2}\cdot\theta\exp(-i\mu_K t)$, 
where $\theta$ is the chiral angle, $f$ (= 93 MeV) the meson decay constant, and $\mu_K$ is the $K^-$ chemical potential. 
The $\sigma$-$K^-$ coupling strength $g_{\sigma K}$ is related to the $K^-$ optical potential depth $U_K$
in symmetric nuclear matter. We set $U_K$=$-$120 MeV as a moderately attractive case. See \cite{4} for details of the other parameters. 

Taking into account the charge neutrality, we construct the effective energy density, ${\cal E^{\rm eff}}\equiv {\cal E}+\mu (\rho_p-\rho_{\Sigma^-}-\rho_{\Xi^-}-\rho_{K^-}-\rho_e)$, where $\mu$ is the charge chemical potential and $\rho_i$ ($i=p, \Sigma^-, \Xi^-, K^-, e^-$)  the number densities. 
The ground state is obtained under the charge neutrality condition, $\partial{\cal E^{\rm eff}}/\partial \mu=0$, and chemical equilibrium conditions for weak processes, 
 which are reduced to $\partial{\cal E^{\rm eff}}/\partial \rho_i=0$ ($i=p, n, \Lambda, \Sigma^-, \Xi^-$) with a given baryon number density $\rho_{\rm B}$. 
 From the last conditions the relations between the chemical potentials, $\mu=\mu_K=\mu_e=\mu_n-\mu_p$, $\mu_\Lambda=\mu_n$, $\mu_{\Sigma^-}=\mu_{\Xi^-}=\mu_n+\mu_e$, are assured.

\section{NUNERICAL RESULTS}

\begin{figure}[h]
\begin{minipage}[l]{0.50\textwidth}
\begin{center}
\includegraphics[height=.30\textheight]{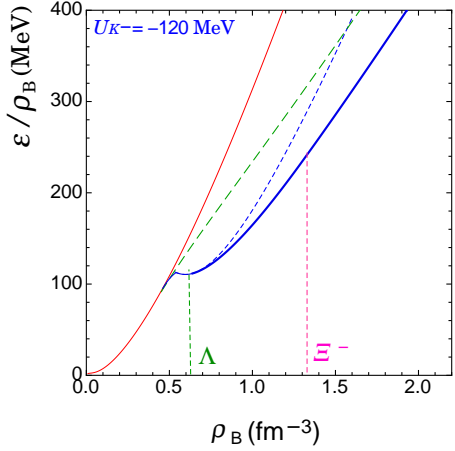}
\end{center}
\caption{The energy per baryon in the ($Y+K$) phase as a function of baryon number density $\rho_{\rm B}$ for $U_K=-120$ MeV (the bold solid line). See the text for details.}
\label{fig1}
\end{minipage}~
\begin{minipage}[r]{0.50\textwidth}
as compared with the case of hyperonic matter without kaon condensates.
 This is due to the fact that the $K^-$-$\Xi^-$ vector interaction works repulsively as far as $\mu>0$, as a consequence of chiral symmetry. 
The $\Sigma^-$ hyperons do not appear over the relevant densities. This is because the $\Sigma^-$ potential depth in nuclear matter is taken to be repulsive, which is deduced from the recent hypernuclear experiments\rf{4}. 

One can see that 
once kaon condensation occurs in hyperonic matter, it leads to significant softening of the EOS,  since the attractive effect of the $s$-wave $K^-$-$B$ interaction is added as well as the effect of avoiding nucleon ($N$)-$N$ repulsion through mixing of hyperons\rf{5}. 
\end{minipage}
\end{figure}

In Fig.~\ref{fig1}, the energy per baryon in the ($Y+K$) phase is shown as a function of $\rho_{\rm B}$ for $U_K=-120$ MeV (the bold solid line). For comparison, those for hyperonic matter without kaon condensation (the long dashed line), for kaon-condensed phase without hyperon-mixing (the dotted line), and for the normal nucleon matter (the thin solid line) are shown. (We found numerical errors in the previous result of the ${\cal E^{\rm eff}}$-$\rho_{\rm B}$ relation presented at the conference. They have been corrected in this proceedings.)  
Kaon condensation sets in  
at $\rho_{\rm B}=2.9\rho_0 $ with $\rho_0$ (=0.153 fm$^{-3}$) being the nuclear saturation density, just before the onset of hyperons. In the kaon-condensed phase, only the $\Lambda$ hyperons start to appear at $\rho_{\rm B}$= 4.2 $\rho_0$, and the fraction of the $\Lambda$  monotonically increases with density. At a higher density $\rho_{\rm B}=8.8 \rho_0$, the $\Xi^-$ hyperons set in. The onset density of the $\Xi^-$ hyperons is pushed up to high density 
 
\section{DISCUSSION AND SUMMARY}

One should take into account stiffening effects on the kaon-condensed EOS in hyperonic matter at high baryon density. For example, many-body forces between baryons such as three-body $NNN$, $YNN$, $YYN$, $YYY$ forces should be considered beyond the mean-field approximation\rf{5}.
It is also instructive to consider anti-symmetrization effect for baryons beyond the present Hartree approximation through introduction of tensor coupling of vector mesons\rf{6}. 
It should be noted that there is another higher energy state consisting of kaon-condensed phase with hyperon-mixing in addition to the ground state one at some density intervals. 
This ${\cal E^{\rm eff}}$-$\rho_{\rm B}$ branch may lead to stiffer EOS than that for the ground state. We will carefully examine whether the existence of the higher energy state 
is specific to this model or universal result stemming from interplay between kaon condensates
and hyperons.


\begin{thebibliography}{999}

\bibitem{1} P.~B.~Demorest et al., Nature~{\bf 467},~108~(2010); J.~Antoniadis et al.,Science~{\bf 340},~6131~(2013).

\bibitem{2} T.~Muto, Phys.~Rev.~{\bf C 77},~015810~(2008), and references cited therein.

\bibitem{3} D.~B.~Kaplan, A.~E.~Nelson, Phys.~Lett.~{\bf B 175},~57~(1986).

\bibitem{4} T.~Muto, T.~Maruyama, T.~Tatsumi, Phys.~Rev.~{\bf C 79},~035207~(2009); {\it ibid}, JPS Conf.~Proc.~{\bf 1}, 013081~(2014).

\bibitem{5} S.~Nishizaki, Y.~Yamamoto, T.~Takatsuka, Prog.~Theor.~Phys.~{\bf 108}, 703~(2002). 

\bibitem{6} T.~Miyatsu, T.~Katayama, K.~Saito, Phys.~Lett.~{\bf B 709},~242~(2012).

\end{thebibliography}
\end{document}